\RequirePackage{amsmath}
\documentclass[sigconf]{acmart}
\usepackage{graphicx}
\usepackage{color}
\usepackage{soul}
\usepackage{verbatim}
\usepackage{subfig}
\usepackage{enumitem}
\usepackage{url,footnote}
\usepackage{dirtytalk}
\usepackage[utf8]{inputenc}
\usepackage[T1]{fontenc}
\usepackage[english]{babel}
\usepackage{sidecap}
\usepackage{wrapfig}
\usepackage{multirow}
\usepackage{listings}
\usepackage{amsmath}
\usepackage{stmaryrd}
\usepackage{amssymb}
\usepackage[linesnumbered,ruled,lined,boxed,commentsnumbered]{algorithm2e}
\usepackage{ifsym}
\usepackage{booktabs}
\usepackage{tikz}
\usepackage{rotating}
\usepackage{todonotes}
\usepackage{flushend}

\definecolor{codegreen}{rgb}{0,0.6,0}
\definecolor{codegray}{rgb}{0.5,0.5,0.5}
\definecolor{codepurple}{rgb}{0.58,0,0.82}
\definecolor{backcolour}{rgb}{0.95,0.95,0.92}
 
\lstdefinestyle{mystyle}{
    backgroundcolor=\color{backcolour},   
    commentstyle=\color{codegreen},
    keywordstyle=\color{magenta},
    numberstyle=\tiny\color{codegray},
    stringstyle=\color{codepurple},
    basicstyle=\footnotesize,
    breakatwhitespace=false,         
    breaklines=true,                 
    captionpos=b,                    
    keepspaces=true,                 
    numbers=left,                    
    numbersep=2pt,                  
    showspaces=false,                
    showstringspaces=false,
    showtabs=false,                  
    tabsize=1
}
 
\hypersetup{
    colorlinks=true,
    linkcolor=blue,
    filecolor=blue,      
    urlcolor=cyan,
}

\begin{document}


\title{Killing Two Birds with One Stone -- Querying Property Graphs using SPARQL via \textsc{Gremlinator}}





\author{Harsh Thakkar}
\affiliation{%
  \institution{University of Bonn}
  \city{Bonn} 
  \country{Germany} 
  \postcode{53117}
}
\email{thakkar@cs.uni-bonn.de}

\author{Dharmen Punjani}
\affiliation{%
 \institution{National and Kapodistrian University of Athens}
 \city{Athens} 
 \country{Greece} 
 \postcode{10679}}
\email{dpunjani@di.uoa.gr}

\author{Jens Lehmann}
\affiliation{
  \institution{University of Bonn \& Fraunhofer IAIS}
  \city{Bonn}
  \country{Germany} 
  \postcode{53117}}
\email{jens.lehmann@cs.uni-bonn.de}

\author{S{\"o}ren Auer}
\affiliation{%
  \institution{TIB \& Leibniz University of Hannover}
  \city{Hannover}
  \country{Germany} 
  \postcode{30167}}
\email{soeren.auer@tib.eu}

\renewcommand{\shortauthors}{H. Thakkar et al.}
\renewcommand{\shorttitle}{Killing Two Birds with One Stone via Gremlinator}

\begin{abstract}
Knowledge graphs have become popular over the past decade and frequently rely on the Resource Description Framework (RDF) or Property Graph (PG) databases as data models.  
However, the query languages for these two data models -- SPARQL for RDF and the PG traversal language Gremlin -- are lacking interoperability.
We present \textbf{Gremlinator}, the first translator from SPARQL -- the W3C standardised language for RDF -- and Gremlin -- a popular property graph traversal language. 
Gremlinator translates SPARQL queries to Gremlin path traversals for executing graph pattern matching queries over graph databases. 
This allows a user, who is well versed in SPARQL, to access and query a wide variety of Graph Data Management Systems (DMSs) avoiding the steep learning curve for adapting to a new Graph Query Language (GQL). 
Gremlin is a graph computing system-agnostic traversal language (covering both OLTP graph database or OLAP graph processors), making it a desirable choice for supporting interoperability for querying Graph DMSs.
Gremlinator currently supports the translation of a subset of SPARQL 1.0, specifically the SPARQL \texttt{SELECT} queries.

\end{abstract}

%


\maketitle

\section{Introduction}
Knowledge graphs model the real world in terms of entities and relations between them. They became popular as they are an intuitive and simple data model, which allows to execute many types of queries efficiently and can serve as a foundation for a range of Artificial Intelligence applications. The Resource Description Framework (RDF) and Property Graphs (PGs) are popular languages for knowledge graphs. For RDF, the SPARQL query language was standardized by W3C, whereas for PGs several languages are frequently used, including Gremlin~\cite{rodriguez2015traversalmachine}.

PGs and RDF have evolved from different origins and still have largely disjoint user communities. RDF is part of the Semantic Web initiative with a focus on expressive data modelling as well as data publication and linking. PGs originate from the database community with a focus on efficient execution of graph traversals.

With \textit{Gremlinator}, we build a bridge between both communities and research the interoperability of RDF and PG query languages~\cite{ThakkarGremlinator2018}. Moreover, we allow combining the best of both worlds: Powerful modelling capabilities as well as data publication and interlinking methods combined with efficient graph traversal execution. In particular, Gremlinator has the following advantages:
(1) Existing SPARQL-based applications can switch to property graphs in a non-intrusive way. 
(2) It provides the foundation for a hybrid use of RDF triple stores and property graph DMS -- a system could detect which DMS is more efficient for answering a particular query~\cite{das2014tale} and redirect the query accordingly. In particular, property graph databases have been shown to work very well for a wide range of queries which benefit from locality in a graph. Rather than performing expensive joins, property graph databases use micro indices to perform traversals. 
(3) Users familiar with the W3C standardized SPARQL query language do not need to learn another query language.

Overall, we make the following contributions:
\begin{itemize}[nosep]
    \item A novel approach for mapping SPARQL queries to Gremlin pattern matching traversals, Gremlinator, which is the first work converting an RDF to a property graph query language to the best of our knowledge.
    \item An openly available implementation for executing SPARQL queries over a plethora of third party graph DMS such as \textit{Neo4J}, \textit{Sparksee}, \textit{OrientDB}, etc. using the \textit{Apache TinkerPop} framework. 
\end{itemize}

The remainder of the article is organized as follows: Section~\ref{sec:rel_work} summarizes the related work. Section~\ref{sec:grem_approach} sheds light on the importance of Gremlin, briefly discusses the Gremlinator approach and its limitations. Section~\ref{sec:demo} presents the demonstration details and the value Gremlinator will cater to its users. Finally, Section~\ref{sec:concl_fw} concludes the article and describes the future work.

\section{Related Work}\label{sec:rel_work}
We present a brief summary of related work with regard to techniques and tools that support the translation and execution of formal query languages, addressing the interoperability issue.

\textbf{SPARQL $\rightarrow$ SQL:} There is a substantial amount of work been done for conversion of SPARQL queries to SQL queries, such as -- \textit{Ontop}~\cite{calvanese2017ontop}, \textit{R2RML}~\cite{rodriguez2015efficient}, Elliot et al.~\cite{elliott2009complete}, Chebotko et al.~\cite{chebotko2009semantics}, Zemke et al.~\cite{zemke2006converting}, Priyanka et al.~\cite{priyatna2014formalisation}. 
\emph{Ontop}~\cite{calvanese2017ontop}, one of the most popular system, exposes relational databases as virtual RDF graphs by linking the terms (classes and properties) in the ontology to the data sources through mappings. This virtual RDF graph can then be queried using SPARQL.

\textbf{SQL $\rightarrow$ SPARQL:} \emph{RETRO}~\cite{rachapalli2011retro} presents a formal semantics preserving the translation from SQL to SPARQL. It follows a schema and query mapping approach rather than to transform the data physically. The schema mapping derives a domain-specific relational schema from RDF data. Query mapping transforms an SQL query over the schema into an equivalent SPARQL query, which in turn is executed against the RDF store.

\textbf{SQL $\rightarrow$ CYPHER:} CYPHER\footnote{CYPHER Query Language~(\url{https://neo4j.com/developer/cypher-query-language/})} is the graph query language used to query the \emph{Neo4j}\footnote{\label{Neo4j}Neo4j~(\url{https://neo4j.com/})} graph database. 
There has been no work yet to use SQL on top of CYPHER. 
However, there are some examples\footnote{SQL to CYPHER~(\url{https://neo4j.com/developer/guide-sql-to-cypher/})} that show the  equivalent CYPHER queries for certain SQL queries.


\section{Gremlinator Approach}\label{sec:grem_approach}
In this section we discuss the why we choose Gremlin as a property graph query language and briefly describe the Gremlinator approach.
\subsection{Why Apache TinkerPop Gremlin?}
Gremlin is a system-agnostic query language developed by Apache TinkerPop\footnote{Gremlin: Apache TinkerPop's graph traversal language and machine~(\url{https://tinkerpop.apache.org/})}. It supports both -- pattern matching (declarative) and graph traversal (imperative) style of querying over property graphs.

\begin{figure}[htbp]
\begin{center}
\includegraphics[width=0.5\textwidth]{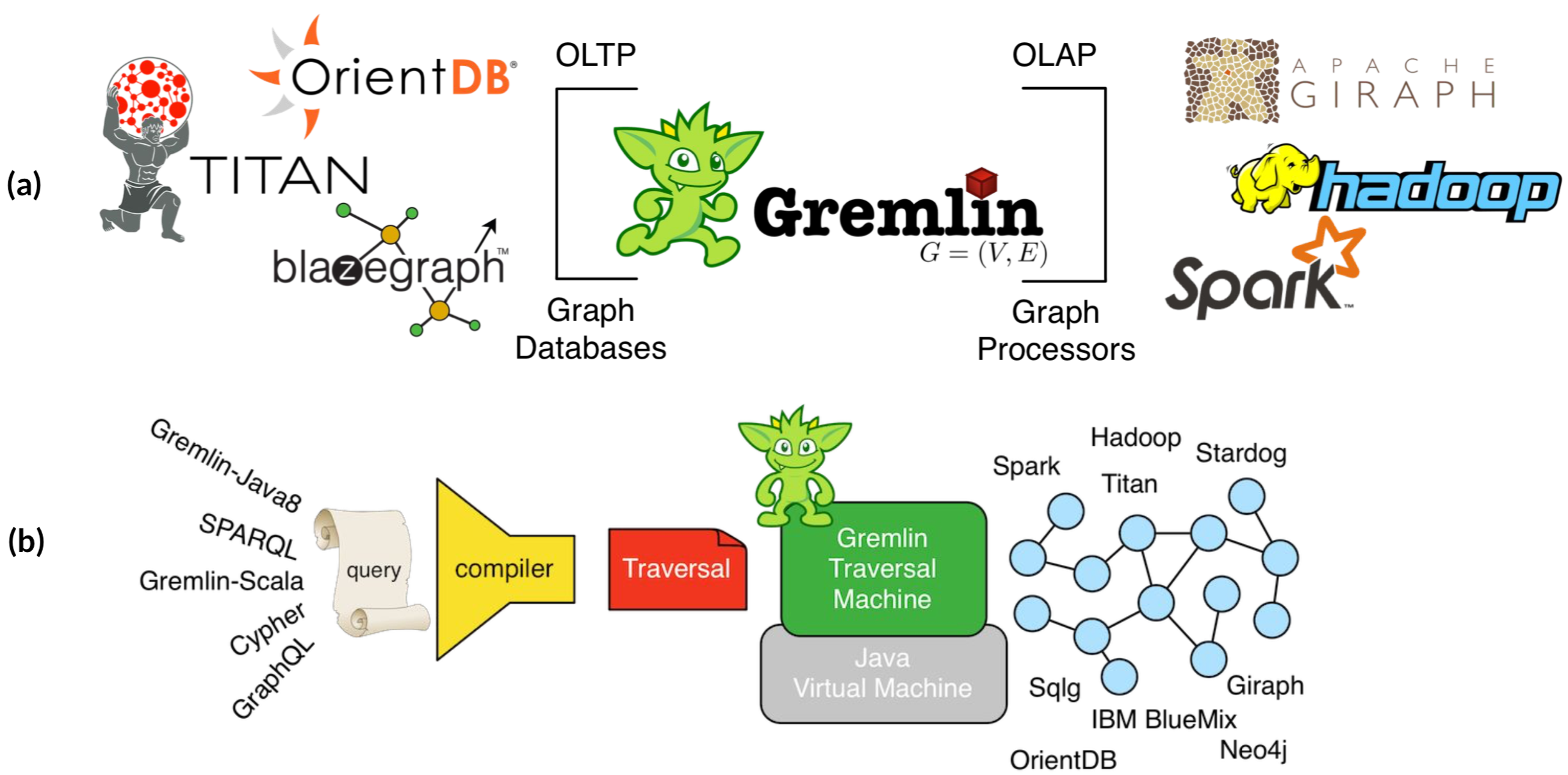}
\end{center} \vspace{-5pt}
\caption{\textbf{The Gremlin Traversal Language and Machine.}}
\label{fig:tinker-all}
\end{figure}
Gremlin is more general than, e.g.~,CYPHER, as it provides in addition to a query language a common execution platform for supporting any graph computing system (including both OLTP and OLAP graph processors), for addressing the querying interoperability issue (see Figure~\ref{fig:tinker-all} (a)).
Together with Apache TinkerPop framework, Gremlin is a language and a virtual machine, it is possible to design another traversal language that compiles to the Gremlin traversal machine (analogous to how Scala compiles to the JVM), ref. Figure~\ref{fig:tinker-all} (b).
Gremlin provides the declarative (SPARQL style) pattern matching querying construct using the \texttt{.match()}-step.

For brevity, we abstain from dwelling into the definitions and formal semantics of Gremlin, rather point the interested reader to the literature~\cite{rodriguez2015traversalmachine,thakkar2017graph}. Furthermore, one may also refer to~\cite{ThakkarGremlinator2018}, where the complete SPARQL to Gremlin translation approach is discussed in detail.

\subsection{Gremlinator Pipeline}
\begin{figure}[htbp]
\begin{center}
\includegraphics[width=0.5\textwidth]{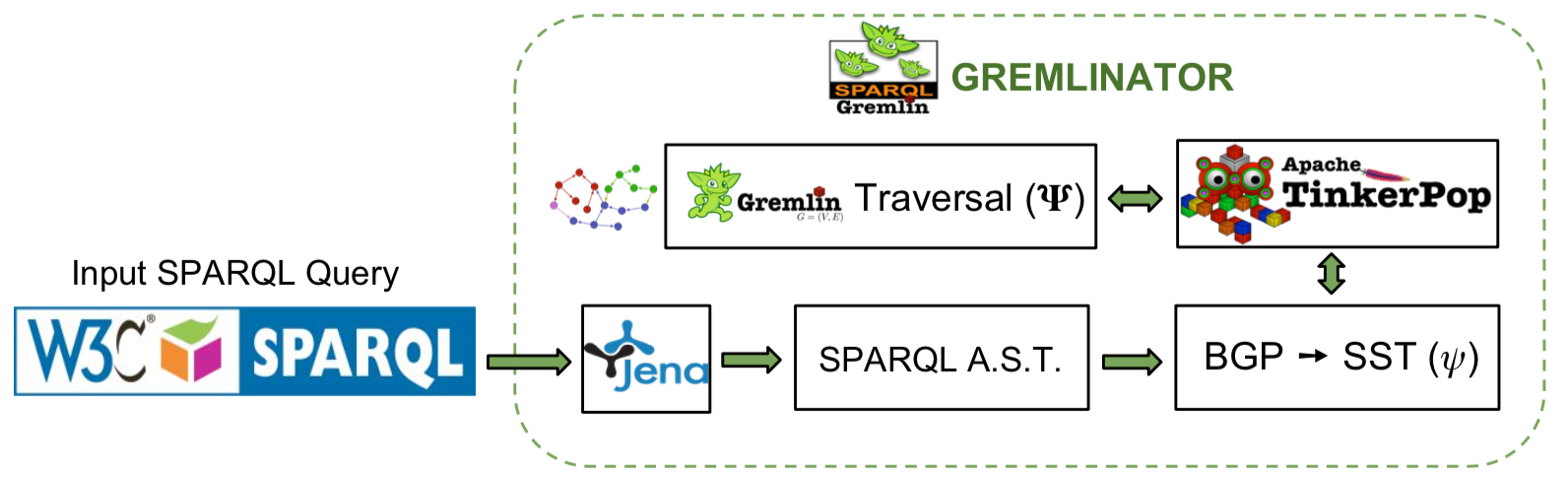}
\end{center} \vspace{-5pt}
\caption{\textbf{The architectural overview of Gremlinator.}}
\label{fig:grem_arch}
\end{figure}

We now present the architectural overview of Gremlinator in Figure~\ref{fig:grem_arch} and discuss each of the four steps of its execution pipeline. 

\textbf{Step 1.} The input SPARQL query is first parsed using the Jena ARQ module, thereby: (i) validating the query and (ii) generating its abstract syntax tree (AST) representation. 

\textbf{Step 2.} From the obtained AST of the parsed SPARQL query, Gremlinator then visits each basic graph pattern (BGP), mapping them to the corresponding Gremlin single step traversals (SSTs). 
A SST in Gremlin is an atomic traversal step ($\psi_{s}$) we describe in~\cite{ThakkarGremlinator2018} in detail.

\textbf{Step 3.} Thereafter, depending on the operator precedence obtained from the AST of the parsed SPARQL query, each of the corresponding SPARQL keywords are mapped to their corresponding instruction steps from the Gremlin instruction library. Thereafter a final conjunctive traversal ($\Psi$) is generated appending the SSTs and instruction steps.
This can be perceived analogous to the SPARQL query language, wherein a set of BGPs form a single complex graph pattern (CGP).

\textbf{Step 4.} 
This final conjunctive traversal ($\Psi$) is used to generate bytecode\footnote{Bytecode is simply serialized representation of a traversal, i.e. a list of ordered instructions where an instruction is a string operator and a (flattened) array of arguments.} which can be used on multiple language and platform variants of the Apache TinkerPop Gremlin family. 

\subsection{Pre-defined Queries}

\textbf{Considerations.} We encode the prefixes of SPARQL queries within the Gremlinator implementation.  In order to aid the SPARQL to Gremlin translation process, we define custom prefixes preserving the categories of Gremlin instruction steps.
For instance, the standard \texttt{rdfs:label} prefix (which is generally a predicate) is represented as \texttt{e:label} or \texttt{v:label} (where e = edge and v =  vertex).

For the demonstration of Gremlinator, we provide a set of 30 pre-defined SPARQL queries for reference, for each dataset, covering 10 different SPARQL query features (i.e. three queries per feature with a combination of various modifiers) as shown in Table~\ref{tab:query_characs}.
These features were selected after a systematic study of SPARQL query semantics~\cite{AnglesFMGQLs16,perez2006semantics,schmidt2010foundations} and from \textit{BSBM}~\cite{bsbm} explore use cases\footnote{BSBM Explore Use Cases~(\url{https://goo.gl/y1ObNN})} and \textit{Watdiv Query templates}\footnote{Watdiv Query Features~(\url{http://dsg.uwaterloo.ca/watdiv/basic-testing.shtml})}.
Furthermore, we encourage the end user to write and execute custom SPARQL queries for both the datasets, for further exploration.

\begin{table}[tb]
\centering
\caption{Query feature and  description}
\label{tab:query_characs} \vspace{-10pt}
\resizebox{0.5\textwidth}{!}{%
\begin{tabular}{lll}
\toprule
\textbf{Query Id.} & \textbf{Feature}  & \textbf{Description} \\
\midrule
C1-C3 & CGPs &  Queries with mixed number of BGPs \\
F1-F3 & FILTER &  CGPs with a combination of $\geq$1 FILTER constraints \\
L1-L3 & LIMIT+OFFSET  & CGPs with a combination of LIMIT + OFFSET constraints \\
G1-G3 & GROUP BY &  CGPs with GROUP BY feature \\
Gc1-Gc3 & GROUP COUNT & CGPs with GROUP BY + COUNT \\
O1-O3 & ORDER BY &  CGPs with ORDER BY feature \\
U1-U3 & UNION &  CGPs with UNION feature \\
Op1-Op3 & OPTIONAL &  CGPs with OPTIONAL BGPs \\
M1-M3 & MIX &  CGPs with a combination of all above features \\
S1-S3 & STAR & CGPs forming a STAR shape execution plan ($\geq$10 BGPs) \\
\bottomrule
\end{tabular}
}
\end{table} 

\subsection{Limitations}
    Gremlinator is an on-going effort for achieving seamless translation of SPARQL queries to Gremlin traversals. 
    The current version of Gremlinator supports the SPARQL 1.0 \texttt{SELECT} queries with the following excceptions: 
    1.) \texttt{REGEX} (regular expressions) in \texttt{FILTER} (restrictions) of a graph pattern are currently not supported.
    2.) Gremlinator does not support variables for the property predicate, i.e.~the predicate \texttt{\{p\}} in a graph pattern \texttt{\{s p o .\}} has to be defined or known for the traversal to be generated. 
    This is because traversing a graph is not possible without knowing the precise traversal operation to the destination (vertex or edge) from the source (vertex or edge).
    
\section{Demonstration Details}\label{sec:demo}

\begin{figure}
    \centering
    \includegraphics[width=0.5\textwidth]{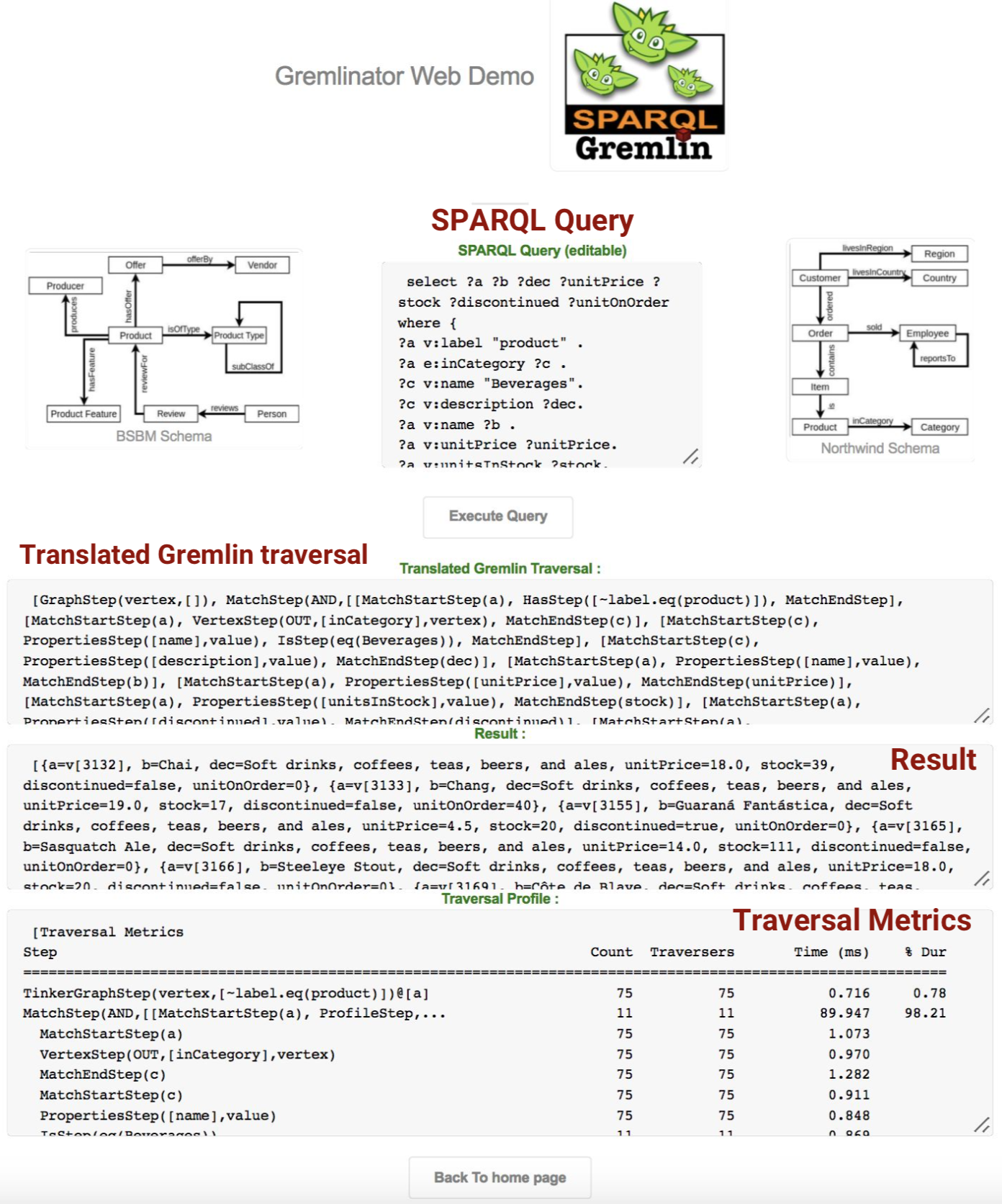}
    \caption{Gremlinator Web application demonstration screenshot.}
    \label{fig:grem_demo}
    \end{figure}

    As a part of the demonstration of our system Gremlinator, we provide-- 
    \textbf{(i)} an online screencast\footnote{Gremlinator Demo Screencast --~\url{https://youtu.be/Z0ETx2IBamw}} 
    \textbf{(ii)} a web application, 
    see Figure~\ref{fig:grem_demo})\footnote{Gremlinator Web Demo --~\url{http://gremlinator.iai.uni-bonn.de:8080/Demo}}
    \textbf{(iii)} a desktop application of Gremlinator (standalone .jar bundle) which requires Java 1.8 JRE installed on the corresponding host machine, downloadable from the web demo website.
    
    The demonstration work-flow for all the above mentioned Gremlinator versions is identical, wherein --
    \textbf{(i)} the user selects a dataset (Northwind or BSBM) from the corresponding drop-down menu;
    \textbf{(ii)} the user selects a query (one of the ten SPARQL query features) from the corresponding drop-down menu;
    \textbf{(iii)} the user executes the query;
    \textbf{(iv)} Gremlinator returns the selected SPARQL query, the translated Gremlin traversal and the result of the traversal execution;
    \textbf{(v)} the user can also edit or write custom SPARQL queries and execute them at selected dataset using the integrated query editor at will.
    
    We will present the live demonstration of Gremlinator using a pre-configured laptop with all the resources including the SPARQL queries and datasets.
    In order to demonstrate the correctness of our approach we will provide a custom docker-based Openlink Virtuoso SPARQL endpoint, pre-loaded with the datasets, for a one-to-one query result comparison (for interested visitors).
    
\begin{figure}[ht]
\begin{center}
\includegraphics[width=0.5\textwidth]{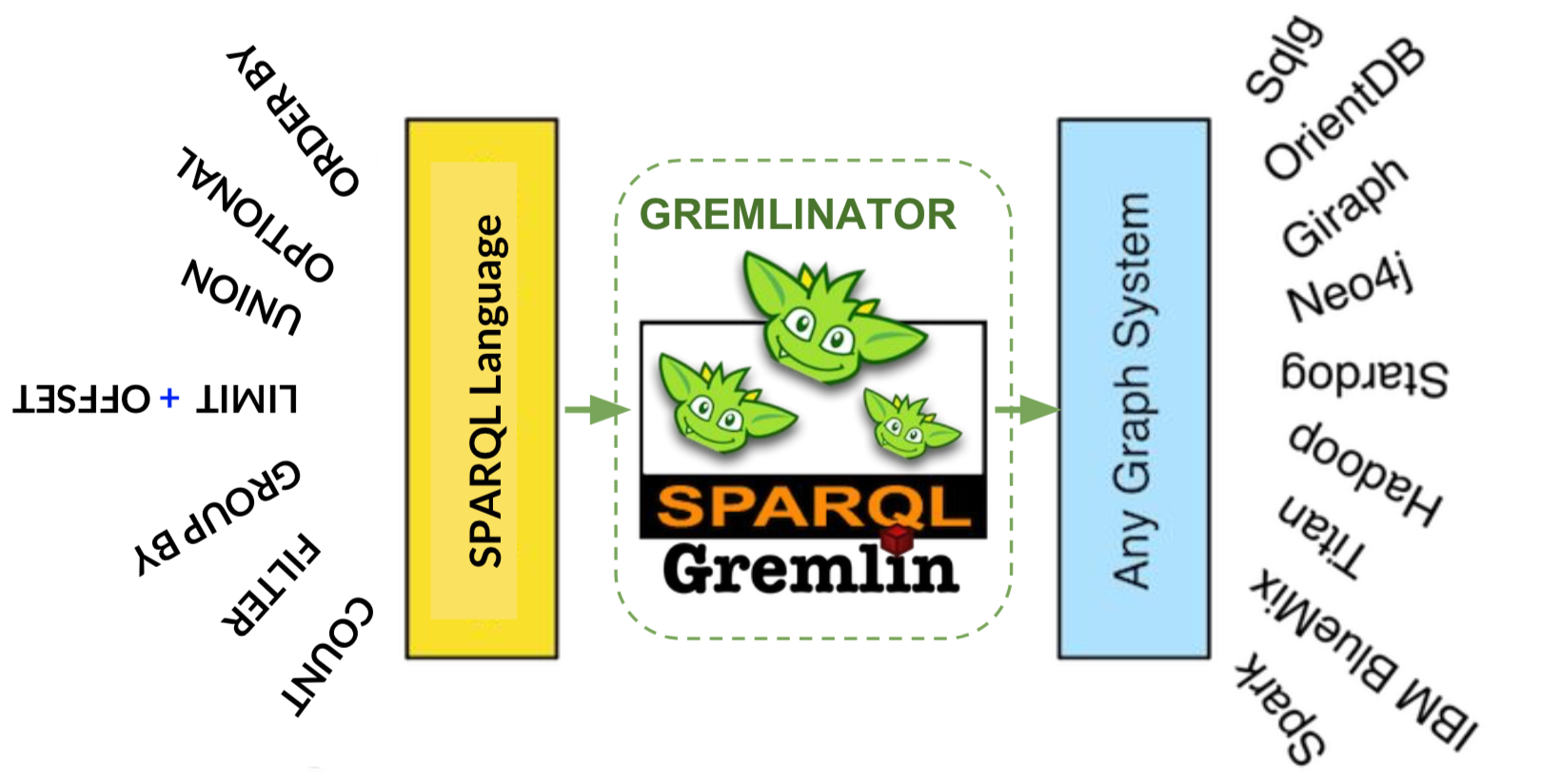}
\end{center} \vspace{-5pt}
\caption{\textbf{Gremlinator powered by Apache TinkerPop will enable querying a variety of Graph databases.}}
\label{fig:grem_bigger}
\end{figure}

    
\textbf{Value.} Gremlinator will serve as a user friendly medium to -- (i) execute SPARQL queries over property graphs bridging the query interoperability gap; (ii) conduct performance analysis of query results, comparisons of SPARQL vs. Gremlin traversal operations using frameworks such as \textit{LITMUS}~\cite{DBLP:conf/esws/Thakkar17,DBLP:conf/i-semantics/ThakkarKDLA17}; and (iii) enable querying a spectrum of graph databases via SPARQL 1.0 query fragment (ref. Figure~\ref{fig:grem_bigger}), leveraging the advantages of the Apache TinkerPop framework.

\section{Conclusion \& Future Work}\label{sec:concl_fw}
We presented a demonstration of Gremlinator, a novel approach for supporting the execution of SPARQL queries on property graphs using Gremlin traversals.
Gremlinator has obtained clearance by the Apache Tinkerpop development team and is currently in production phase to be released as a plugin during TinkerPop's next framework cycle. Gremlinator has also been integrated into the \href{http://sansa-stack.net/}{SANSA Stack}~\cite{iswc_sansa} (v0.3) framework as an experimental plugin. 
Furthermore, Gremlinator is freely available under the Apache 2.0 license for public use from the \href{https://mvnrepository.com/artifact/io.github.litmus-benchmark-suite/gremlinator}{Maven Central} repository. 

As future work, we are working on -- 
(i) adding support for \texttt{REGEX} in restriction (FILTERs), variables for property predicates, and (ii) supporting translation of SPARQL 1.1 query features such as property paths, in the upcoming releases.
    
\section{Author Biographies}
    \vspace{-15pt}
    \begin{table}[H]
    \begin{center}
    \begin{tabular}{ c p{6.4cm} }
    \raisebox{-\totalheight}{\includegraphics[trim={5cm 0 2cm 0},clip,width=20mm,scale=0.5]{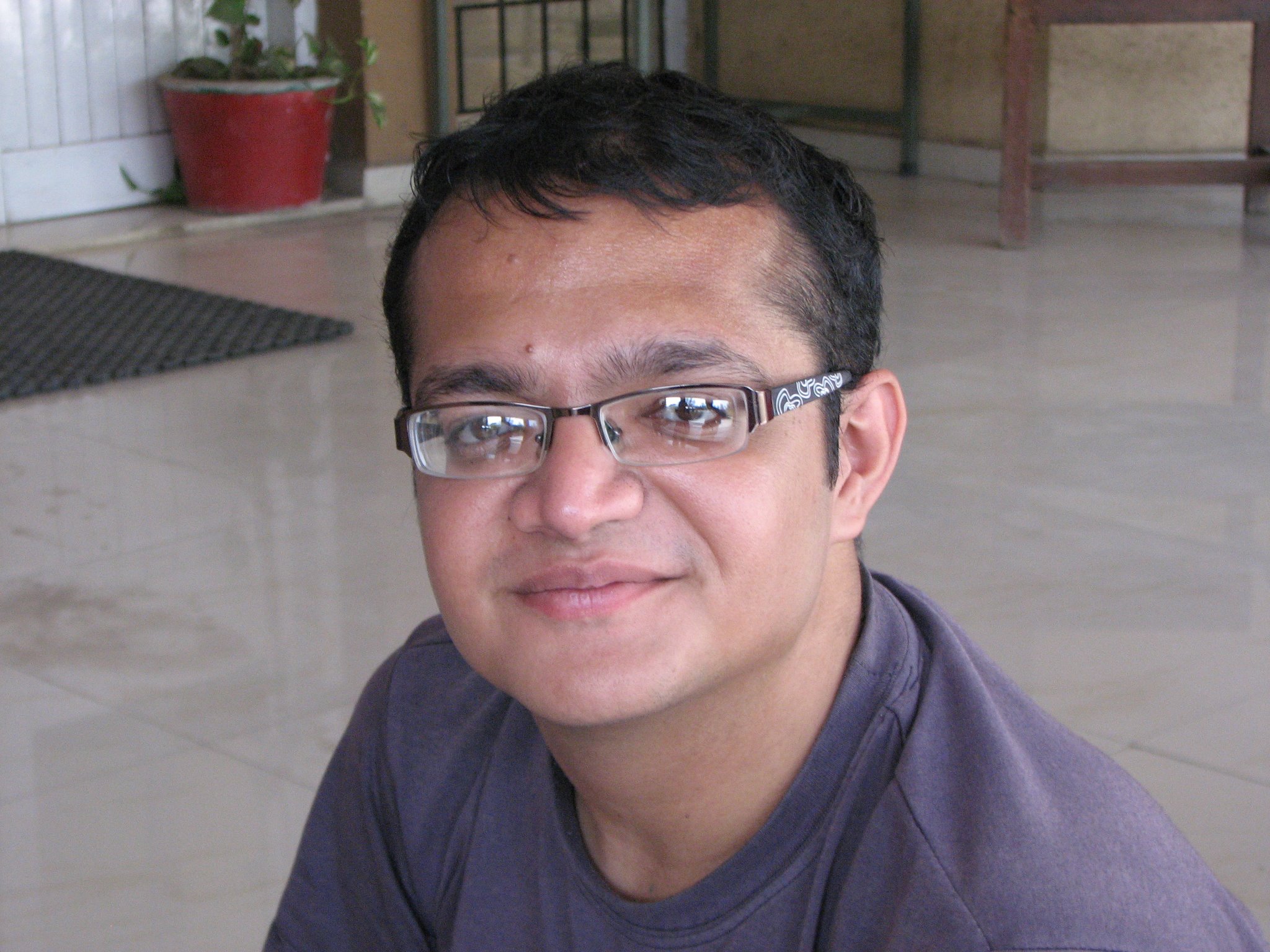}}
    & \textbf{Harsh Thakkar} - is a Marie Sk\l{}odowska-Curie Ph.D. student at the University of Bonn, Germany. He earned his M.Tech. in Computer Science from NIT Surat, India. 
    His research interests include Graph and RDF Data Management, Benchmarking, Graph Query Languages and Question Answering.
    \\ 
    \raisebox{-\totalheight}{\includegraphics[trim={6cm 0 0 5cm},clip,width=20mm,scale=0.5]{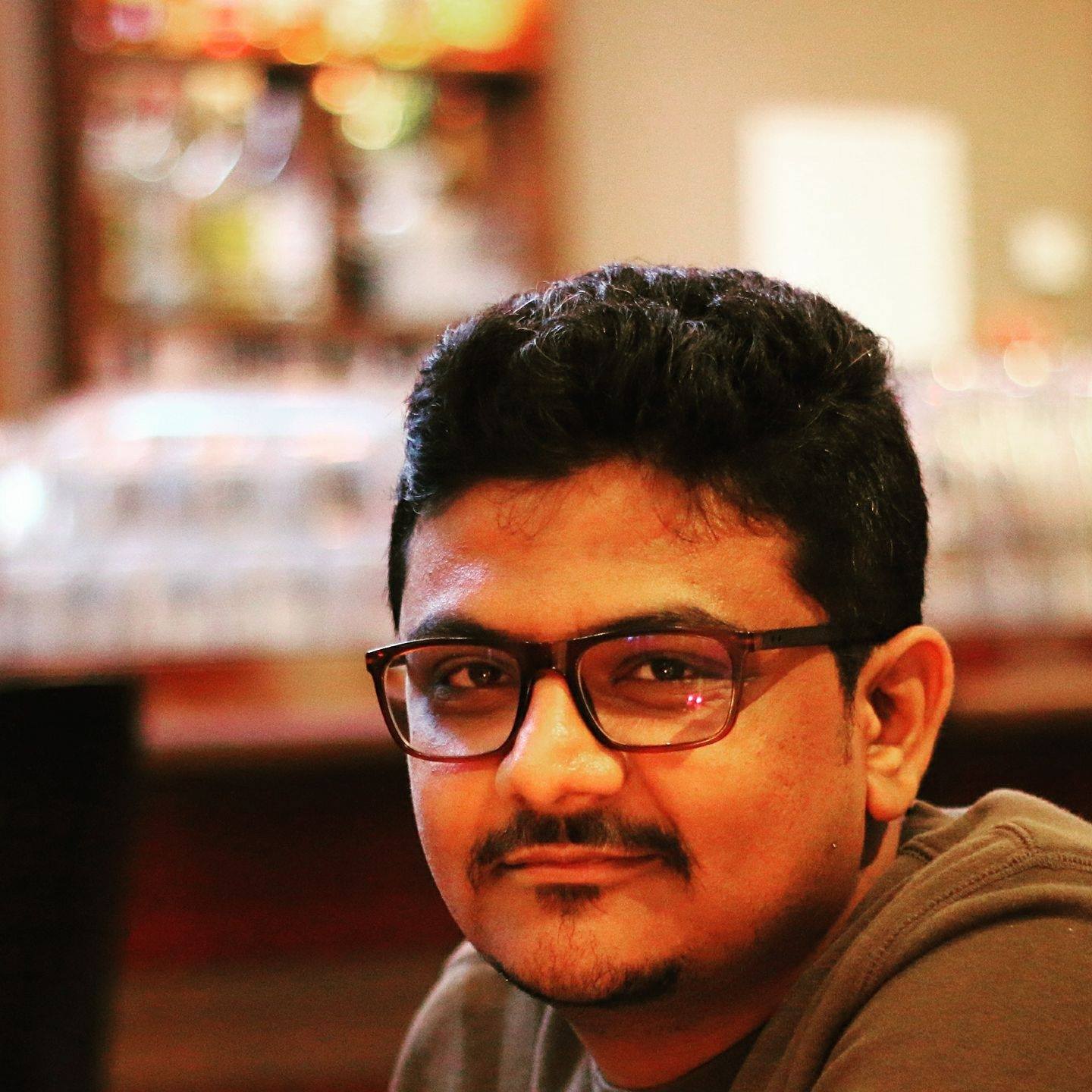}}
    & \textbf{Dharmen Punjani} - is a Marie Sk\l{}odowska-Curie Ph.D. student at the National and Kapodistrian University of Athens, Greece. He earned his M.Tech. in Computer Science from NIT Surat, India. 
    His research interests include Geo-Spatial and RDF Data Management, N.L.P., and Question Answering.
    \\
    \raisebox{-\totalheight}{\includegraphics[trim={0 1cm 0 0.3cm},clip,width=20mm,scale=0.5]{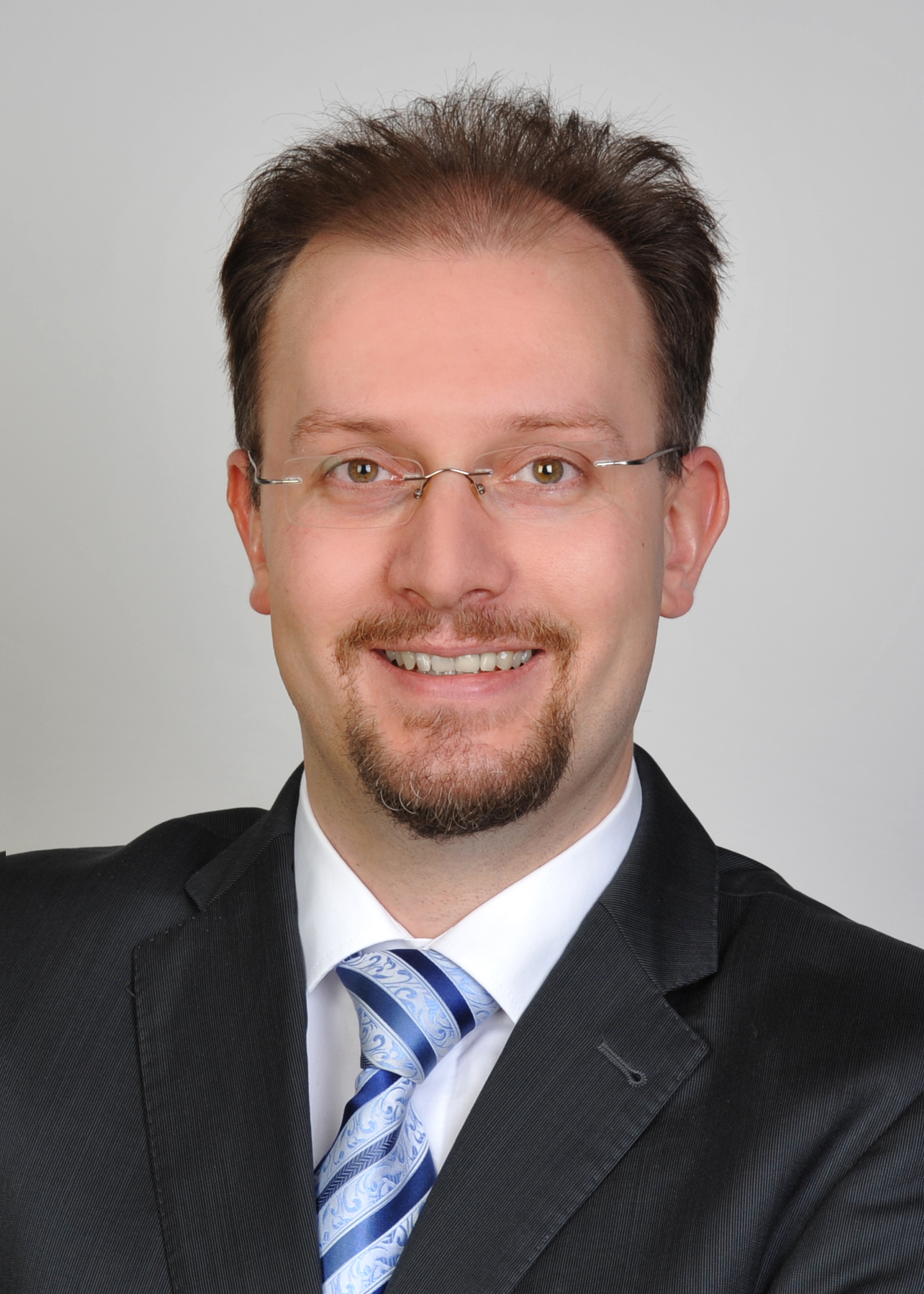}}
    & \textbf{Jens Lehmann} is professor for Software and Data Engineering, leads the Smart Data Analytics (SDA) research group at the University of Bonn and is a lead scientist at the Enterprise Information Systems (EIS) department at Fraunhofer IAIS. His main research interests are semantic technologies and machine learning.
    \\
    \raisebox{-\totalheight}{\includegraphics[width=20mm,scale=0.5]{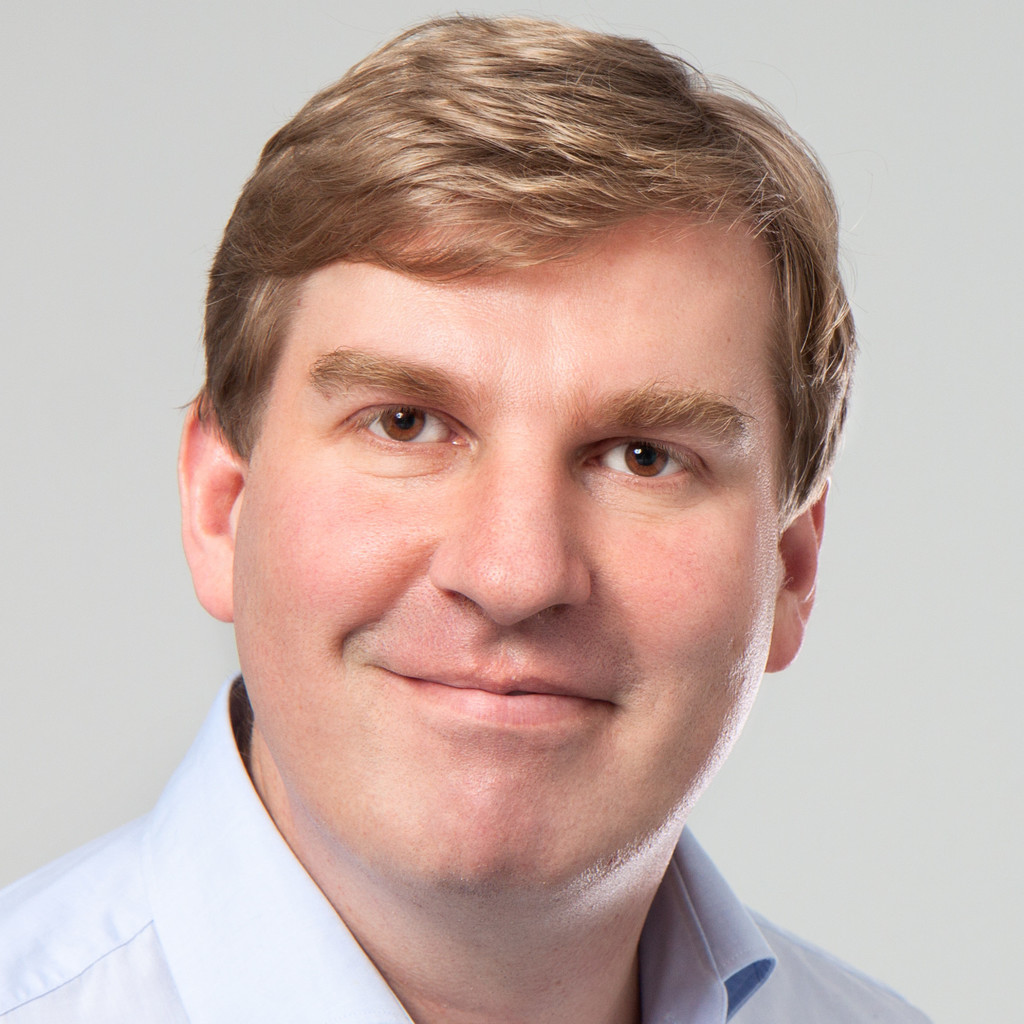}}
    & \textbf{S\"oren Auer} is professor for Data Science and Digital Libraries at University of Hannover and director of TIB Leibniz Information Center for Science and Technology. His research interests revolve around semantic technologies, scholarly communication and digital libraries.
    
    \end{tabular}
    \label{tbl:auth_bib}
    \end{center}
    \end{table}
\section*{Acknowledgements}
This work is supported by the funding received from EU-H2020 \href{http://wdaqua.eu/}{WDAqua ITN} (GA. 642795). We would like to thank Dr. Marko Rodriguez and Mr. Daniel Kuppitz, of the Apache TinkerPop project, for their support and quality insights in developing Gremlinator.

\bibliographystyle{ACM-Reference-Format}
\bibliography{ref} 

\end{document}